\documentclass[runningheads,a4paper]{llncs}

\usepackage{amssymb}
\usepackage{graphicx}
\usepackage{url}
\usepackage{amsmath}
\usepackage{algorithm}
\usepackage{algpseudocode}
\usepackage{multirow}

\newcommand{\vect}[1]{\boldsymbol{#1}} 

\urldef{\mailsa}\path|{alfred.hofmann, ursula.barth, ingrid.haas, frank.holzwarth,|
\urldef{\mailsb}\path|anna.kramer, leonie.kunz, christine.reiss, nicole.sator,|
\urldef{\mailsc}\path|erika.siebert-cole, peter.strasser, lncs}@springer.com|
\newcommand{\keywords}[1]{\par\addvspace\baselineskip
\noindent\keywordname\enspace\ignorespaces#1}

\begin{document}

\mainmatter  

\title{Development of Krylov and AMG linear solvers for large-scale sparse matrices on GPUs}

\titlerunning{Lecture Notes in Computer Science: Authors' Instructions}

\author{Bo Yang, Hui Liu, Zhangxin Chen}
\authorrunning{Lecture Notes in Computer Science: Authors' Instructions}

\institute{Department of Chemical and Petroleum Engineering\\
University of Calgary\\
Calgary, Canada\\
\{yang6, hui.j.liu, zhachen\}@ucalgary.ca}

\toctitle{Lecture Notes in Computer Science}
\tocauthor{Krylov and AMG linear solvers on GPUs}
\maketitle

\begin{abstract}
This research introduce our work on developing Krylov subspace and AMG solvers on NVIDIA GPUs.
As SpMV is a crucial part for these iterative methods, SpMV algorithms for single GPU and multiple GPUs are implemented.
A HEC matrix format and a communication mechanism are established. And also, a set of specific algorithms for solving preconditioned
systems in parallel environments are designed, including ILU(k), RAS and parallel triangular solvers.
Based on these work, several Krylov solvers and AMG solvers are developed.
According to numerical experiments, favorable acceleration performance is acquired
from our Krylov solver and AMG solver under various parameter conditions.
\end{abstract}

\section{Introduction}
Iterative algorithms have widely applications in kinds of scientific computing fields, such as the reservoir simulation \cite{john}. For large-scale sparse linear systems, Krylov subspace and AMG algorithms are commonly used. Krylov subspace algorithms include the GMRES (Generalized Minimal Residual), CG(Conjugate Gradient) and BiCGSTAB (Biconjugate Gradient Stabilized), etc. These algorithms are available to general matrices \cite{saad,template}. Preconditioners are always employed to optimize the performance of an iterative algorithm and many efficient proecondtioners have been developed \cite{jcxu,CHM,g_mstage,hui}. We have developed the Krylov subspace algorithms with ILU preconditioners. Many researchers have devoted their efforts into designing AMG solvers which is specific for symmetric positive definite matrices. Ruge and St\"{u}ben designed the {RS (Ruge-St\"{u}ben) coarsening strategy and developed a classical AMG solver  which is the foundation of developing other AMG solvers \cite{Stuben1,Stuben2,BMR,CW,PV}. The parallel coarsening strategy CLJP was proposed by Luby, Jones and Plassmann \cite{hypre,rsamg}. We have also developed the AMG algorithm with a series of smoothers, coarsening operators and prolongation operators.

GPU(Graphics Processing Unit) computing emerges as an acceleration technique for image displaying. However, it has more and more utility in other scientific computing disciplines. Zhang et at. completed some professional performance analysis about GPUs \cite{pzhang2}. A NVIDIA Tesla K40 which has 2880 CUDA cores and a peak performance of 1.43 TFlops (Base Clocks) in double precision has greater performance than
an Intel Core i7-5960X with 8 cores and 16 threads which has a typical peak performance of 385 GFlops \cite{k80k40,fujitsu5960x}. A NVIDIA Tesla K40 also has 288 G/sec memory speed which is much faster than the speed 68 GB/s of an Intel Core i7-5960X \cite{k80k40,fujitsu5960x}. As GPU has great priority in parallel computing, we have designed and developed our iterative algorithms on GPUs.

SpMV (sparse matrix-vector multiplication) is a core part for iterative algorithms. For a large and sparse matrix, it is necessary to partition it into sub matrices for GPU computation. The METIS partition method is adopted in our algorithms \cite{metis}.  Because data communication is unavoidable for SpMV implementation on multiple GPUs, we have designed a specific communication mechanism for partition matrices to share vector data among different GPUs. In order to make full use of the characteristic of GPU memory access, we adopted a HEC matrix format which is more friendly to the SpMV algorithm. A NVIDIA GPU platform provides high parallel capability depending on its hundreds of fine CUDA cores. An algorithm must be designed as a parallel algorithm to run on the CUDA cores. RAS (Restricted Additive Schwarz) proposed by Cai et at. is adopted in our algorithms to improve the parallel structure of a preconditioner matrix \cite{cai}. Because the ILU preconditioners and AMG smoothers all need to solve triangular systems, we implemented a parallel triangular solver on GPUs \cite{spmv-uc}. It is based on the level schedule method \cite{saad,LS_gpil}. In this research, we designed a set of numerical experiments to test our algorithms from different aspects. The experiment results and analysis are given in the experiment section.

The layout of this paper is presented as follows:
In \S \ref{sec_gpu_computation}, the matrix format, SpMV, vector operations, ILU (k), RAS, parallel triangular solver, Krylov subspace algorithms and AMG algorithms are introduced.
In \S \ref{sec_experiment}, the numerical experiments are presented and analyzed.
In \S \ref{sec_conclusion}, conclusions are given.

\section{GPU Computation}
\label{sec_gpu_computation}

\subsection{Matrix Format}
Several matrix formats are presented in this section. They are ELL, HYB and HEC. The ELL format is provided in ELLPACK \cite{ellpack}. Figure~\ref{fig_ell} shows the ELL's structure consisting of two parts. We can see the two parts are both regular and have the same dimensions. Regular storage has a high speed for data access. However, it is not wise to store a large-scale sparse matrix in such a format as lots of storage spaces are always wasted. For instance, if there are a large number of nonzero entries in one row, the other rows must maintain the same size of entries most of which are zero. In order to make the limited memory space be used efficiently, N. Bell and M. Garland suggested a hybrid matrix format named HYB (Hybrid of ELL and COO). An original matrix is split into two parts. One part is regular and the remain part is irregular. The COO format is used to store the irregular part. It has three one-dimensional arrays illustrated in Figure~\ref{fig_coo}. The HYB format has good average performance. In our research, we adopt another hybrid format called HEC which saved the irregular part in a CSR format shown in Figure~\ref{fig_csr}. It also contains three one-dimensional arrays. $Ap$ is used for storing the start position of each row. $Ax$ and $Aj$ have the same length and used for storing the entry data and column indices, respectively.

\begin{figure}[!tbh]
    \centering
    \includegraphics[width=0.6\linewidth]{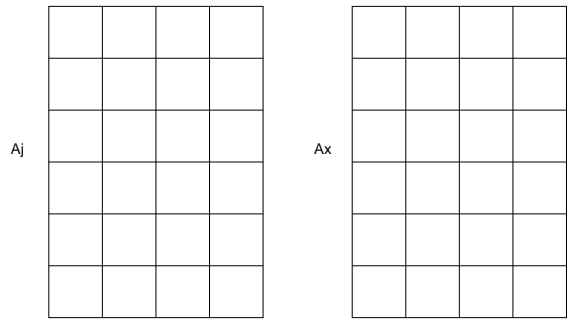}
    \caption{ELL matrix format}
    \label{fig_ell}
\end{figure}

\begin{figure}[!tbh]
    \centering
    \includegraphics[width=0.6\linewidth]{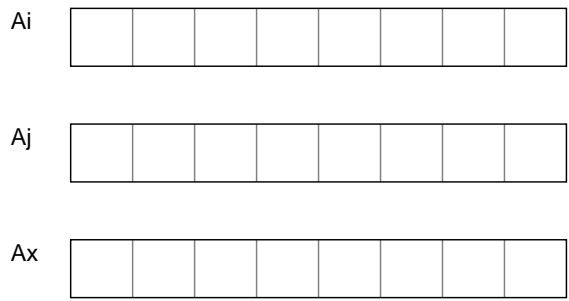}
    \caption{COO matrix format}
    \label{fig_coo}
\end{figure}

\begin{figure}[!tbh]
    \centering
    \includegraphics[width=0.6\linewidth]{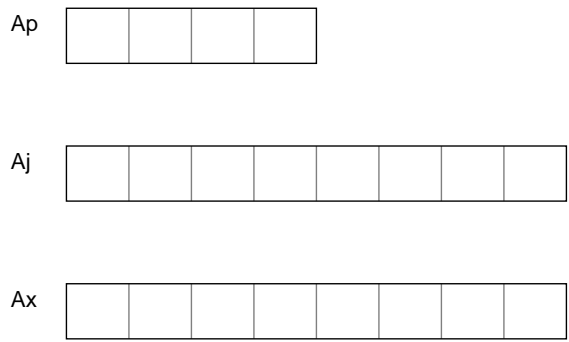}
    \caption{CSR matrix format}
    \label{fig_csr}
\end{figure}

According to the mathematic method of SpMV, it is always calculated based on the column vectors. This can be explained by equation \eqref{equ_spmv_c}. Thereby, it's better for us to store the entries in the computer column by column. The GPU architecture provides a wrap concept to execute CUDA cores. That means 32 threads are bounded to be executed together. So the stride of the ELL part should be a multiple of 32 to acquire enhanced parallel performance. In our algorithms, we set it as 256 or other multiples. Another problem is how to decide the boundary between the ELL and CSR. We use a recommended value 20 whose theoretical explanation are introduced in \cite{nv-spmv2}.

\begin{equation}
\label{equ_spmv_c}
\left(
  \begin{array}{cccc}
    A_{11}  &A_{12}  &\cdots  &A_{1n}\\
    A_{21}  &A_{22}  &\cdots  &A_{2n}\\
    \vdots  &\vdots  &\ddots  &\vdots\\
    A_{n1}  &A_{n2}  &\cdots  &A_{nn}\\
  \end{array}
\right)
\left(
  \begin{array}{c}
    x_1\\
    x_2\\
    \vdots\\
    x_n\\
  \end{array}
\right)
=
x_1\left(
  \begin{array}{c}
    A_{11}\\
    A_{21}\\
    \vdots\\
    A_{n1}\\
  \end{array}
\right)
+
x_2\left(
  \begin{array}{c}
    A_{12}\\
    A_{22}\\
    \vdots\\
    A_{n2}\\
  \end{array}
\right)
+
\cdots
+
x_n\left(
  \begin{array}{c}
    A_{1n}\\
    A_{2n}\\
    \vdots\\
    A_{nn}\\
  \end{array}
\right)
\end{equation}

\subsection{SpMV Algorithm}
Based on the HEC matrix format, the SpMV algorithm is designed as two parts apparently. As a GPU executes hundreds of CUDA cores simultaneously, a parallel algorithm can be implemented with each CUDA core computing a row. The ELL part has high efficient and is performed firstly. Algorithm~\ref{alg_spmv_single} gives the SpMV algorithm. This algorithm runs well on a single GPU. However, it is not suitable for multiple GPUs. Multiple GPUs bring stronger parallel computing capability but import extra data communication. We need to partition the original matrix into partition matrices first.

\begin{algorithm}[htb]
\caption{Sparse matrix-vector multiplication}
\label{alg_spmv_single}
\begin{algorithmic}[1]
\For {i = 1: n} \Comment{ELL}
  \State Calculate the $i$-th row of ELL matrix; \Comment{one CUDA core}
\EndFor
\State
\For {i = 1: n} \Comment{CSR}
  \State Calculate the $i$-th row of CSR matrix; \Comment{one CUDA core}
\EndFor
\end{algorithmic}
\end{algorithm}

\begin{figure}[!tbh]
    \centering
    \includegraphics[width=0.6\linewidth]{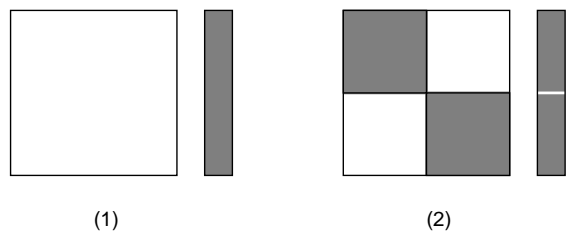}
    \caption{Matrix and vector partition}
    \label{fig_partition}
\end{figure}

If a matrix has a regular structure. For instance, it is derived from the FDM (Finite Difference Method) or FVM (Finite Volume Method). A sequence partition method can be used. But if it is irregular structure which is often derived from the FEM (Finite Element Method) or FVM. A specific partition method should be used. We select a quasi-optimal partition method METIS to complete the matrix partition. During the partition process, the rows of the matrix are switched first and all the nonzero entries are put along the diagonal as close as possible. Then the pivot blocks have most of the nonzero entries and the communication cost between any two partition matrices is reduced; see Figure~\ref{fig_partition}.

\begin{figure}[!tbh]
    \centering
    \includegraphics[width=0.8\linewidth]{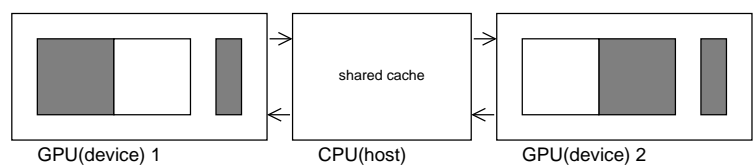}
    \caption{Vector communication}
    \label{fig_communication}
\end{figure}

The vector is also partitioned into segments. Each pair of a partition matrix and a segment vector is distributed onto a GPU. Although most of the nonzero entries are concentrated at the pivot block, there are still some sparse nonzero entries outside it, for which a segment vector can not provide a corresponding element to complete multiplication. Thus, the necessary communication is unavoidable. We establish a shared cache for communication. The cache is located on the CPU (host). It receives all the communication data from each GPU (device) and then sends the data to needed GPU. As we have used partition method to reduce the communication load, this mechanism is reasonable.

\subsection{Vector Operations}

Vector operations are necessary for developing iterative algorithms. They can be categorized into some categories by equation~\eqref{e_vector_updates_1} to equation~\eqref{e_norm}. Some of them are linear combinations of vectors. Some of them are about dot products. As two vectors are operated by one-to-one correspondence of elements, it is easy to design parallel algorithms for them. First, vectors are divided into segments. Then each pair of segments are distributed onto a GPU. All the sub results are sent back to CPU after tasks are finished on GPUs. No communication cost is needed during the vector operations. A schematic is shown by Figure~\ref{fig_vectoroperations}.

\begin{equation}
\label{e_vector_updates_1}
\vect{y} = \alpha A \vect{x} + \beta \vect{y}
\end{equation}

\begin{equation}
\label{e_vector_updates_2}
\vect{y} = \alpha \vect{x} + \beta \vect{y}
\end{equation}

\begin{equation}
\label{e_vector_updates_3}
\vect{z} = \alpha \vect{x} + \beta \vect{y}
\end{equation}

\begin{equation}
\label{e_dot_product}
a = \langle \vect{x}, \vect{y} \rangle
\end{equation}

\begin{equation}
\label{e_norm}
r = \|\vect{x}\|_2 = \sqrt{\langle \vect{x}, \vect{x}\rangle}
\end{equation}

\begin{figure}[!tbh]
    \centering
    \includegraphics[width=0.8\linewidth]{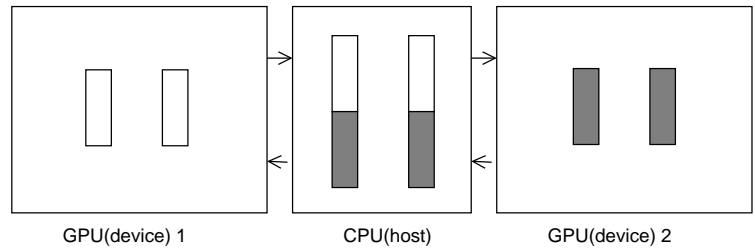}
    \caption{Vector operations}
    \label{fig_vectoroperations}
\end{figure}

\subsection{ILU(k)}
A preconditioner system is expressed as equation~\eqref{equ_pc}. $M$ is the preconditioner matrix which is factorized from the original matrix $A$. The ILU is a commonly used precondtioner. It means $M$ can be factorized into one lower triangular matrix $L$ and an upper triangular matrix $U$, as shown by equation~\eqref{equ_lu}. The matrix $A$ and $LU$ are stored in the same memory space in the program implementation. In other words, $L$ is stored in the low triangular part and $U$ is stored in the upper triangular part. A level $k$ can be used to control the factorization process. Only the entry positions meeting the requirement are allowed to have nonzero entries in the result pattern. The requirement condition is described by equation~\eqref{equ_L_1} and equation~\eqref{equ_L_2} \cite{saad}.

\begin{equation}
\label{equ_pc}
M\vect{x} = \vect{y}
\end{equation}
where
\begin{itemize}\itemsep1pt \parskip0pt \parsep0pt
  \item $M$: the preconditioner matrix
  \item $\vect{x}$: the unknown vector
  \item $\vect{y}$: the right hand side vector
\end{itemize}

\begin{equation}
\label{equ_lu}
M = LU
\end{equation}

\begin{equation}
\label{equ_L_1}
L_{ij} = \left\{
 \begin{aligned}
 & 0, & (i, j) \in P \\
 & \infty, & (i, j) \notin P.
 \end{aligned}
  \right.
\end{equation}

\begin{equation}
\label{equ_L_2}
L_{ij} = min \{L_{ij}, L_{ip} + L_{pj} + 1\}.
\end{equation}

Equation~\ref{equ_L_1} gives an initial level for each entry $A_{ij}$. $P$ is the nonzero pattern of $A$. So if $A_{ij}$ is zero, its level $L_{ij}$ is infinite; otherwise, $L_{ij}$ is zero. Equation~\eqref{equ_L_1} provides an updated algorithm for levels. This update process are executed at each loop of ILU(k) algorithm and only the satisfactory entry positions have nonzero values in the final factorization pattern. The Algorithm~\ref{alg_pwiluk} details a complete ILU(k) procedure.

\begin{algorithm}[!htb]
\caption{ILU(k) factorization} \label{alg_pwiluk}
\begin{algorithmic}[1]
\State For all nonzero entries in nonzero pattern $P$, define $L_{ij} = 0$
 \For {$i = 2: n$}
  \For {$p = 1: i - 1 $ \& $L_{ip} \le k$}
   \State $A_{ip} = A_{ip} / A_{pp}$
     \For {$j = p + 1: n $}
     \State $A_{ij} = A_{ij} - A_{ip}A_{pj}$
     \State $L_{ij} = min \{L_{ij}, L_{ip} + L_{pj} + 1\}$
     \EndFor
  \EndFor
  \If {$L_{ij} > k$}
     \State $A_{ij} = 0$
  \EndIf
\EndFor
\end{algorithmic}
\end{algorithm}

\subsection{Restricted Additive Schwarz}
A preconditioner system is always solved at least once in a loop of an iterative algorithm. Its solution speed has great influence on the entire solution process. A GPU platform provides hundreds of CUDA cores to complete a parallel task. If we can improve the parallel structure of a preconditioner matrix, the solution process can be accelerated. Cai et al. proposed a Restricted Additive Schwarz method to optimize the parallel structure for a preconditioner, as illustrated by Figure~\ref{fig_RAS}. The original matrix $A$ is partitioned into some sub matrices first. By the METIS method mentioned above, we got these rectangular matrices whose pivot blocks are dense and other positions are sparse; see Figure~\ref{fig_RAS}-(2). Because the ILU factorization only needs an approximate factorization result from $A$, we can remove the sparse entries situated outside the pivot blocks. Analyzed from a graph aspect, the entries in the pivot blocks represent vertices and the entries outside them represent edges. If we remove the edges from the graph by RAS process, the communication among GPUs are ruled out. The remained pivot blocks can be solved in parallel. The improvement of parallel performance leads to an accuracy decrease as we discard some entries. So more iteration times are required to reach a convergence. There is an alternative way named overlap to compensate for the loss of accuracy. As shown by Figure~\ref{fig_RAS}-(3), the overlap technique requires each pivot block to include its some layers of neighbor entries into the block matrices to be computed. Extra entries improve the calculation accuracy and reduce the iteration times. But extra entries also have a negative influence on the parallel performance. Parallelization and convergence like a cake. We cannot eat it and have it. This characteristic is reflected in the numerical experiment section. As a multiple-GPU platform has two levels of parallelization, the situation becomes complex. One level is composed by the GPUs. The other level is the CUDA cores on each GPU. Both levels need a partition and a overlap.

\begin{figure}[!tbh]
    \centering
    \includegraphics[width=0.8\linewidth]{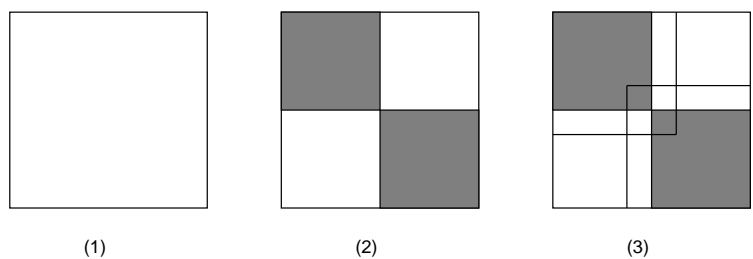}
    \caption{Restricted Additive Schwarz}
    \label{fig_RAS}
\end{figure}
\subsection{Parallel Triangular Solver}

In order to solve $L$ and $U$ on GPUs, we design a parallel triangular solver based on the level schedule method. As an upper triangular system can be easily changed into a lower one, only the lower triangular system is analyzed. The algorithm of a parallel triangular solver is divided into two steps. Each unknown $x(i)$ is assigned a level which is defined by equation~\eqref{eqn_pts_level} in the first step \cite{saad}. The second step is the solution process. The triangular problem is solved level by level. All the unknowns in the same level are solved simultaneously. The first level is dependence free. So it is solved at the very first. After the unknowns in the first level is obtained, the second level becomes free and can be solved. This procedure proceeded until all the levels are computed and all the unknowns are solved. A complete algorithm of the level schedule method is given by Algorithm~\ref{alg_lvs}.

\begin{equation}
\label{eqn_pts_level}
 l(i) = 1 + \max_j {l(j)} \quad \text{ for  all } j \text{ such
that } \ L_{ij} \neq 0, i = 1, 2, \ldots, n,
\end{equation}
where
\begin{itemize}\itemsep1pt \parskip0pt \parsep0pt
  \item $L_{ij}$: the $(i, j)$th entry of $L$
  \item $l(i)$: initialized by zeroes
  \item $n$: the number of rows
\end{itemize}

\begin{algorithm}
\caption{Level schedule method for a lower triangular system, $L\vect{x} = \vect{b}$}
\label{alg_lvs}
\begin{algorithmic}[1]
\State Maximal level is n
\For {k = 1 : n}
  \State start = level(k);
  \State end = level(k + 1) - 1;
  \For {i = start: end}
  \State solve the $i$th row;
  \EndFor
\EndFor

\end{algorithmic}
\end{algorithm}

\subsection{Krylov Iterative Algorithms}
By now, we have explained the SpMV, vector operations, precondtioner systems and parallel solution process. All these are components of an iterative algorithm. Krylov subspace algorithms contain a series of iterative Algorithms, such as CG (Conjugate Gradient), GMRES (Generalized Minimal Residual), BiCGSTAB (Biconjugate Gradient Stabilized), etc. We have implemented all of them. For instance, an implementation analysis of the BiCGSTAB is shown in the Algorithm~\ref{alg_bicgstab}. All the operations on GPUs are commented. Detailed principle of BiCGSTAB and other Krylov subspace algorithms can be found in \cite{saad,template}.

\begin{algorithm}[!htb]
\caption{BiCGSTAB algorithm}
\label{alg_bicgstab}
\begin{algorithmic}[1]

\State $\vect{r_0}  = \vect{b} - A\vect{x_0}$; $\vect{x_0}$ is an initial guess vector                              \Comment{SpMV; Vector update}
\For {k = 1, 2, $\cdots$}
    \State $\rho_{k-1}=(\vect{r_0},\vect{r})$                                                                       \Comment{Dot product}
    \If {$\rho_{k-1} = 0$}
        \State Fails
    \EndIf
    \If {$k=1$}
        \State $\vect{p} = \vect{r}$
    \Else
        \State $\beta_{k-1} = (\rho_{k-1}/\rho_{k-2})(\alpha_{k-1}/\omega_{k-1})$
        \State $\vect{p} = \vect{r} + \beta_{k-1}(\vect{p} - \omega_{k-1}\vect{v})$                                 \Comment{Vector update}
    \EndIf
    \State Solve $\vect{p}^*$ from $M\vect{p}^* = \vect{p}$                                                         \Comment{Preconditioner system}
    \State $\vect{v} = A\vect{p}^*$                                                                                 \Comment{SpMV}
    \State $\alpha_{k} = \rho_{k-1}/(\vect{r}_0, \vect{v})$                                                         \Comment{Dot product}
    \State $\vect{s} = \vect{r} - \alpha_k\vect{v}$                                                                 \Comment{Vector update}
    \If {$\| \vect{s} \|_2$ is satisfied }                                                                          \Comment{Dot product}
        \State $\vect{x} = \vect{x} + \alpha_k\vect{p}^*$                                                           \Comment{Vector update}
        \State Stop
    \EndIf
    \State Solve $\vect{s}^*$ from $M\vect{s}^* = \vect{s}$                                                         \Comment{Preconditioner system}
    \State $\vect{t} = A\vect{s}^*$                                                                                 \Comment{SpMV}
    \State $\omega_k = (\vect{t}, \vect{s})/\| \vect{t} \|^2$                                                       \Comment{Dot product}
    \State $\vect{x} = \vect{x} + \alpha_k\vect{p}^* + \omega_k\vect{s}^*$                                          \Comment{Vector update}
    \State $\vect{r} = \vect{s} - \omega_k\vect{t}$                                                                 \Comment{Vector update}
    \If {$\| \vect{r} \|_2$ is satisfied or $\omega_k = 0$}                                                         \Comment{Dot product}
        \State Stop
    \EndIf
\EndFor

\end{algorithmic}
\end{algorithm}

\subsection{AMG Algorithms}

If the coefficient matrix of a system to be solved is symmetric positive definite, an AMG solver should be a better choice. An AMG algorithm has a $L + 1$ levels architecture. The grid of the level is finer with a smaller level number. So the level 0 is the finest level but the level $L$ is the coarsest level. Figure~\ref{fig_amg} shows the level structure of an AMG solver. An AMG algorithm can be designed as V-cycle, W-cycle or F-cycle. Figure~\ref{fig_amg} is a V-cycle which has the best acceleration effect on a parallel platform. W-cycle has the worst effect. An AMG process has two phases. The first one is called a setup phase in which the coarser grids, the smoothers, the restriction and prolongation operators are all established. The second one is the solution phase in which the multiple-levels system is solved. As a coarser grid has much smaller dimension size compared to its neighbor finer grid, a problem on a coarser grid is easier to be solved. A restriction operation is used for transferring the problem from a finer level to a coarser level. After the problem on the coarser grid is solved, a prolongation operator is used to transfer the solution back to a finer grid. On level $l$, let $A_l$ be the system matrix, $R_l$ be the restriction operator and $P_l$ be the prolongation operator. $S_l$ is the pre-smoother and $T_l$ is the post-smoother. An example AMG algorithm for V-cycle can be designed as Algorithm~\ref{alg_amg}.

\begin{figure}[tbh]
    \centering
    \includegraphics[width=0.70\linewidth]{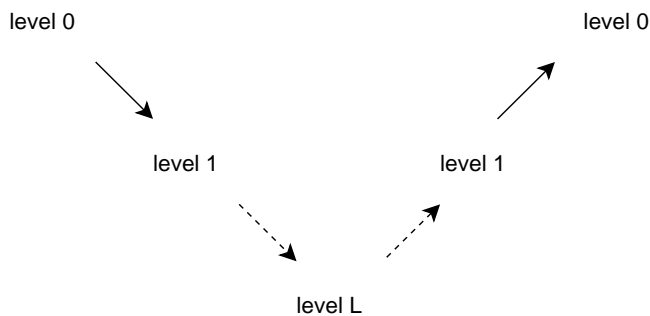}
    \caption{Structure of AMG solver.}
    \label{fig_amg}
\end{figure}

\begin{algorithm}
\caption{AMG V-cycle}
\label{alg_amg}
\begin{algorithmic}
\State Require: $0 \leq l < L$
\State

\If {($l < L$)}
  \State $\vect{x_l} = S_l(\vect{x}_l, A_l, \vect{b}_l)$                                \Comment{Pre-smoothing}
  \State $\vect{r} = \vect{b}_l - A_l\vect{x}_l$
  \State ${\vect{b}_{l+1}} = R_l\vect{r}$                                               \Comment{Restriction}
  \State amg\_solve($l+1$)                                                              \Comment{Recursion}
  \State $\vect{x}_l = \vect{x}_l + P_l \vect{x}_{l+1}$                                 \Comment{Prolongation}
  \State $\vect{x}_l = T_l(\vect{x}_l, A_l, \vect{b}_l)$                                \Comment{Post-smoothing}
\Else
  \State $\vect{x}_l = A_l^{-1}\vect{b}_l$
\EndIf

\end{algorithmic}
\end{algorithm}

We have developed the AMG solver with a series of smoothers, coarsening operators and prolongation operators. The smoothers include damped Jacobi and weighted Jacobi, etc. The coarsening operator RS and the prolongation operator RSSTD are proposed by Ruge and St\"{u}ben  \cite{Stuben1,Stuben2}. The CLJP coarsening operator is proposed by Cleary et al. \cite{hypre,rsamg}.
\section{Numerical Experiments}
\label{sec_experiment}
 A series of numerical experiments are designed to test our algorithms. We use the speedup to measure the parallel acceleration on GPUs. It is calculated by the ratio of the CPU sequential running time to the GPU parallel running time of the same algorithm. The development environment parameters are listed in Table~\ref{tbl_ep}.

\begin{table}[!htb]
\centering
\caption{Experiment environment parameters}
\begin{tabular}{|c|c|} \hline
\bfseries Parameter & \bfseries Value \\ \hline
CPU                             & Intel Xeon X5570          \\ \hline
GPU                             & NVIDIA Tesla C2050/C2070  \\ \hline
Operating System                & CentOS X86\_64            \\ \hline
CUDA Toolkit                    & 5.1                       \\ \hline
GCC                             & 4.4                       \\ \hline
CPU codes compilation           & -O3 option                \\ \hline
float point number precision    & double                    \\ \hline
\end{tabular}
\label{tbl_ep}
\end{table}

\subsection{SpMV}

Table~\ref{tbl_spmv_matrix} gives the properties of matrices used for SpMV test. $3D\_Poisson$ is from a three-dimensional Poisson equation. Its dimension is $150\times150\times150$. The other matrices are all downloaded from a matrix market provided by the University of Florida \cite{mmarket}.

\begin{table}[!htb]
\centering
\caption{Matrices for SPMV}
\begin{tabular}{|c|r|r|r|r|} \hline
\bfseries Matrix    & \bfseries \# of Rows  & \bfseries Nonzeros    & \bfseries NNZ/N & \bfseries  Mb(CSR)\\ \hline
ESOC	           & 327,062	    &6,019,939	&18	       &70\\ \hline
af\_shell8	       & 504,855	    &9,042,005	&18	       &105\\ \hline
tmt\_sym	       & 726,713	    &2,903,837	&4	       &36\\ \hline
ecology2	       & 999,999	    &2,997,995	&3	       &38\\ \hline
thermal2	       & 1,228,045	    &4,904,179	&4	       &61\\ \hline
Hook\_1498	       & 1,498,023	    &30,436,237	&20	       &354\\ \hline
G3\_circuit	       & 1,585,478	    &4,623,152	&3	       &59\\ \hline
kkt\_power	       & 2,063,494	    &7,209,692	&3	       &90\\ \hline
memchip	           & 2,707,524	    &13,343,948	&5	       &163\\ \hline
3D\_Poisson	       & 3,375,000	    &23,490,000	&7	       &282\\ \hline
Freescale1	       & 3,428,755	    &17,052,626	&5	       &208\\ \hline
cage15	           & 5,154,859	    &99,199,551	&19	       &1155\\ \hline
\end{tabular}
\label{tbl_spmv_matrix}
\end{table}

The speedup of SpMV on a single GPU is collected in Table~\ref{tbl_double_single}. Three matrix formats are tested for each matrix. We can see that most of the speedup for HEC format are over 10 and the highest speedup can reach 18. The algorithm on GPUs has good parallel acceleration performance. Figure~\ref{fig_spmv} makes a comparison of different matrix formats. The number of nonzero entries per row is written in the brackets after each matrix name. We can see that the HEC format represented by the red curve shows better performance than the other two formats. From the Figure, the matrices with relative larger $NNZ/N$ have a lower speedup.

\begin{table}[!htb]
\centering
\caption{SPMV speedup for different matrix formats}
\begin{tabular}{|c|r|r|r|} \hline
\bfseries Matrix    &\bfseries  ELL &\bfseries  HYB &\bfseries  HEC \\ \hline
ESOC	       &13.08	  &13.16	      &13.16\\ \hline
af\_shell8	   &9.05	  &10.08	      &11.20\\ \hline
tmt\_sym	   &16.23	  &16.27	      &16.14\\ \hline
ecology2	   &18.38	  &18.24	      &18.11\\ \hline
thermal2	   &8.45	  &8.00	          &9.25\\ \hline
Hook\_1498	   &5.44	  &7.35	          &7.79\\ \hline
G3\_circuit	   &12.84	  &14.08	      &11.22\\ \hline
kkt\_power	   &2.49	  &5.71	          &6.27\\ \hline
memchip	       &4.39	  &10.53	      &11.46\\ \hline
3D\_Poisson	   &13.60	  &13.63	      &13.63\\ \hline
Freescale1	   &5.00	  &9.76	          &11.25\\ \hline
cage15	       &6.40	  &10.00	      &9.89\\ \hline
\end{tabular}
\label{tbl_double_single}
\end{table}

\begin{figure}[!tbh]
    \centering
    \includegraphics[width=0.8\linewidth]{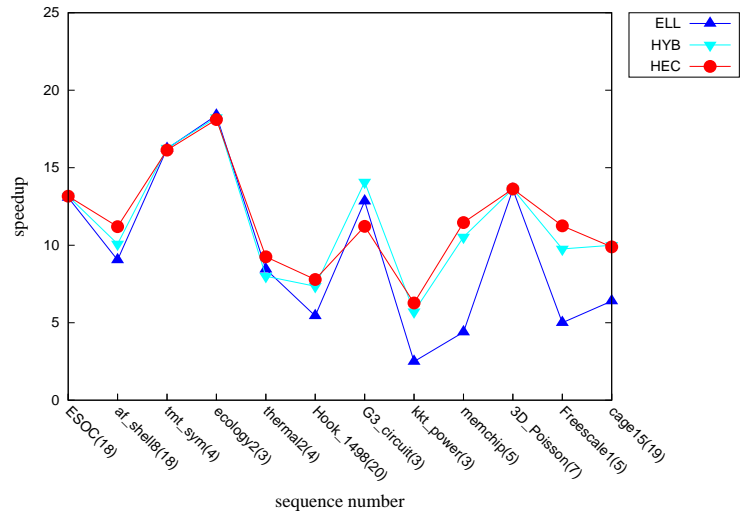}
    \caption{SpMV speedup curves}
    \label{fig_spmv}
\end{figure}

\subsection{BiCGSTAB with ILU(K)}

In this experiment, we use the BiCGSTAB algorithm with an ILU(k) preconditioner to test our Krylov algorithms. The testing matrix is from a three-dimensional Poisson equation whose dimension is 3,375,000 ($150\times150\times150$). It has 23,490,000 nonzero entries and about 7 nonzero entries per row. Table~\ref{table-gpus} collects the running results. There are six parameter combinations which are numbered in the $Seq\ No.$ column. The $Outer\ RAS$ and $Inner\ RAS$ represent the outer layer partition numbers and inner layer partition numbers based on the RAS technique. The number of GPUs employed is equal to the $Outer\ RAS$. The outer and inner overlap layers are listed in the $Outer\ overlap$ and $Inner\ overlap$ columns, respectively. These parameters form various parameter combinations.

\begin{table}[!htb]
\centering
\caption{GMRES with ILU(k) for \texttt{3D\_Poisson} (RAS)}
\begin{tabular}{|r|r|r|r|r|r|r|r|r|r|} \hline
\bfseries Seq & \bfseries Outer & \bfseries Inner & \bfseries Outer & \bfseries Inner & \bfseries ILU(k) & \bfseries CPU time  & \bfseries GPU time  & \bfseries        & \bfseries  \\
\bfseries No. & \bfseries RAS & \bfseries RAS & \bfseries overlap & \bfseries overlap & \bfseries level k & \bfseries (second)	& \bfseries (second)  & \bfseries Speedup  & \bfseries Iteration\\ \hline
1	&1	&8	&0	&0	&0	&16.36	&1.82	&8.97	&45\\
	&	&	&	&	&1	&12.25	&1.56	&7.86	&30\\
	&	&	&	&	&2	&15.75	&3.95	&3.99	&42\\
	&	&	&	&	&3	&16.26	&3.42	&4.75	&33\\
\hline
2	&2	&8	&0	&0	&0	&15.29	&1.07	&14.30	&46\\
	&	&	&	&	&1	&14.64	&1.18	&12.41	&36\\
	&	&	&	&	&2	&18.55	&2.66	&6.96	&43\\
	&	&	&	&	&3	&16.94	&2.80	&6.05	&36\\
\hline
3	&3	&8	&0	&0	&0	&16.57	&0.82	&20.28	&46\\
	&	&	&	&	&1	&14.51	&1.07	&13.59	&39\\
	&	&	&	&	&2	&18.32	&2.53	&7.25	&44\\
	&	&	&	&	&3	&17.85	&2.66	&6.71	&38\\
\hline
4	&4	&8	&0	&0	&0	&17.13	&0.62	&27.84	&44\\
	&	&	&	&	&1	&13.92	&0.81	&17.14	&34\\
	&	&	&	&	&2	&18.15	&2.05	&8.87	&39\\
	&	&	&	&	&3	&17.51	&2.47	&7.08	&38\\
\hline
5	&4	&128	&0	&0	&0	&16.59	&0.62	&26.96	&48\\
	&	&	&	&	&1	&16.91	&0.66	&25.62	&40\\
	&	&	&	&	&2	&20.53	&1.50	&13.72	&51\\
	&	&	&	&	&3	&20.36	&1.56	&13.02	&45\\
\hline
6	&4	&1024	&0	&0	&0	&18.98	&0.67	&28.33	&55\\
	&	&	&	&	&1	&19.37	&0.72	&27.03	&47\\
	&	&	&	&	&2	&21.77	&1.39	&15.63	&58\\
	&	&	&	&	&3	&22.47	&1.27	&17.74	&46\\
\hline
\end{tabular}
\label{table-gpus}
\end{table}

As all the data sections have a similar data tendency, we take the first data section as an sample analysis, where the outer RAS, the inner RAS, the ouer overlap and the inner overlap are 1, 8, 0 and 0, respectively. The speedup reaches 8.97 when the $k$ level is set to 0. As $k$ goes up from 0 to 3, the speedup goes down from 8.97 to 4.75 in a general data tendency. That is because more fill-in entries are imported by a higher $k$. These entries contribute to improve the calculation accuracy. So the iteration is saved and goes down from 45 to 33. However, it goes back to 42 when $k$ is 2. That might is caused by the matrix pattern which has also great influence on the performance.

Figure~\ref{fig_3dpoisson_speedup} shows a comparison of the combinations. As the outer RAS increases from 1 to 4 and then the inner RAS increases from 8 to 1024, the parallel performance is improved gradually and the curves have a growth tendency. Obviously, a lower $k$ has a better parallel performance. The convergence performance is reflected by the Figure~\ref{fig_3dpoisson_iteration}. With the sequence number increases, the parallel performance increases but the convergence performance decreases. Thereby more iteration times are needed to reach a convergence. We can see that high iteration times are needed for $k=0$ because it has high speedup.

\begin{figure}[!tbh]
    \centering
    \includegraphics[width=0.7\linewidth]{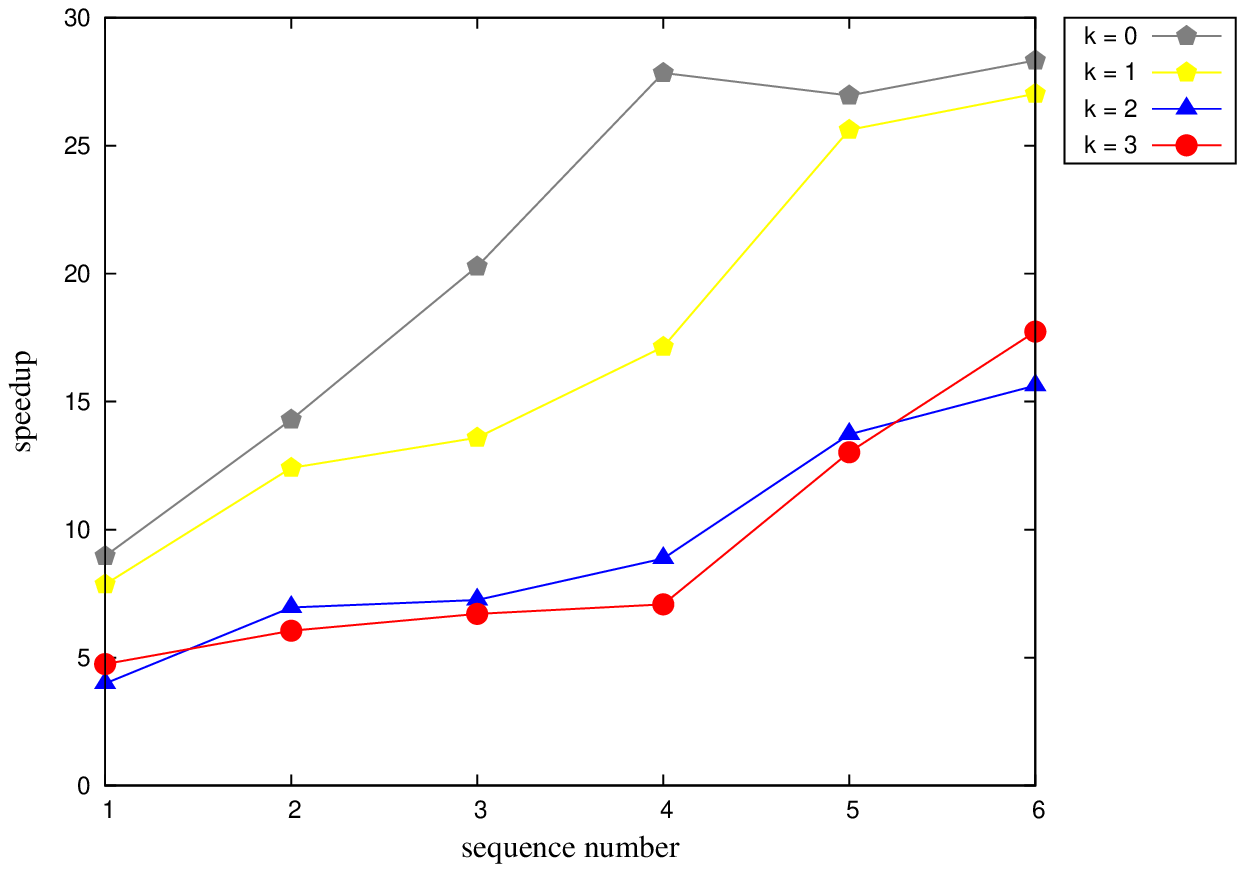}
    \caption{Speedup for \texttt{3D\_Poisson}}
    \label{fig_3dpoisson_speedup}
\end{figure}

\begin{figure}[!tbh]
    \centering
    \includegraphics[width=0.7\linewidth]{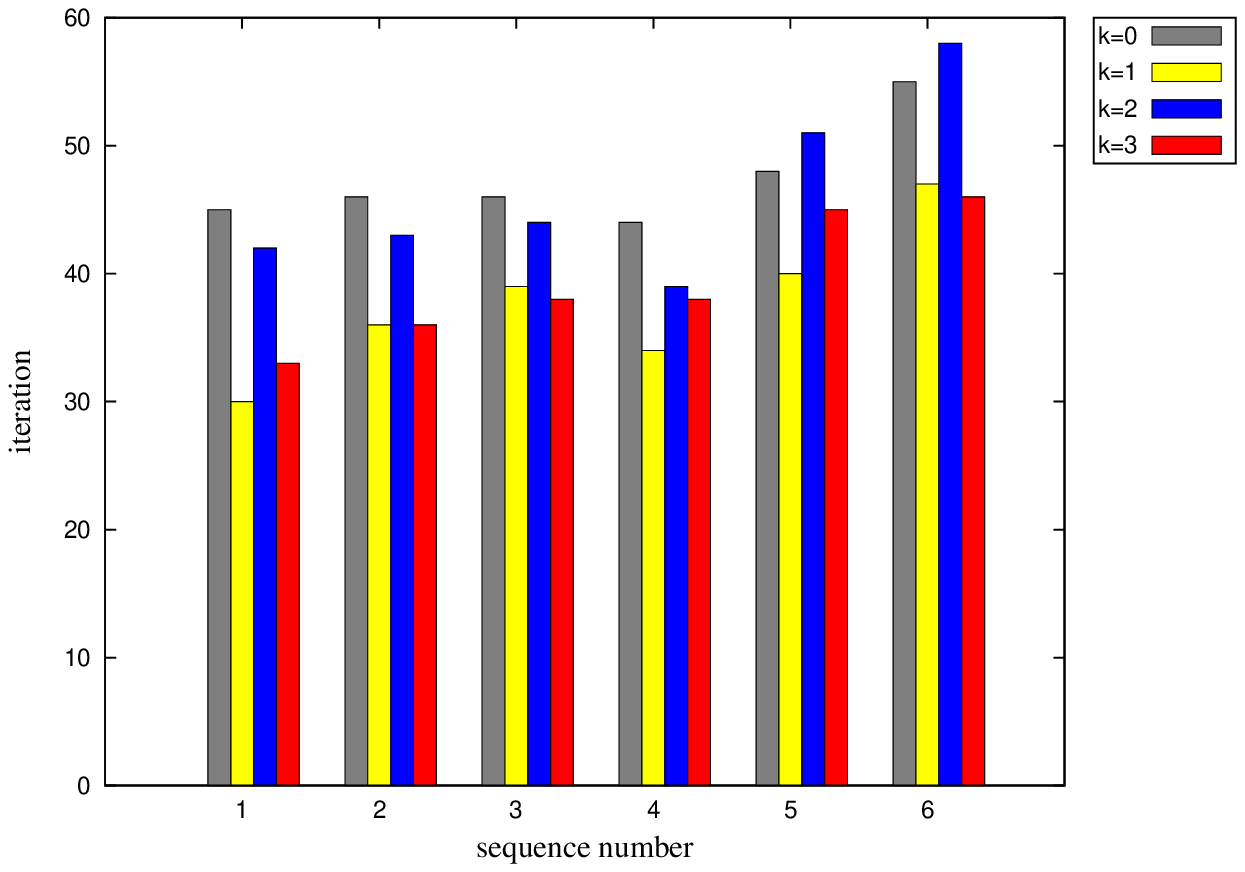}
    \caption{Iteration for \texttt{3D\_Poisson}}
    \label{fig_3dpoisson_iteration}
\end{figure}

As we mentioned, the overlap technique is used for compensating for the loss of calculation accuracy. Higher overlaps are supposed to use smaller iteration. But the speedup is supposed to decrease as more entries are introduced by the overlap. The results of different overlapping configurations are collected in Table~\ref{tbl_overlap}.

\begin{table}[!htb]
\centering
\caption{GMRES with ILU(k) for \texttt{3D\_Poisson} (overlap)}
\begin{tabular}{|r|r|r|r|r|r|r|r|r|r|r|} \hline
\bfseries Seq & \bfseries Outer & \bfseries Inner & \bfseries Outer & \bfseries Inner & \bfseries ILU(k) & \bfseries CPU time  & \bfseries GPU time  & \bfseries        & \bfseries  \\
\bfseries No. & \bfseries RAS   & \bfseries RAS & \bfseries overlap & \bfseries overlap & \bfseries level k & \bfseries (second)	& \bfseries (second)  & \bfseries Speedup  & \bfseries Iteration\\ \hline
1	&4	&8	&0	&0	&0	&17.07	&0.61	&27.82	&44\\ \hline
2	&4	&8	&1	&0	&0	&15.91	&0.70	&22.70	&43\\ \hline
3	&4	&8	&0	&1	&0	&15.43	&0.60	&25.78	&41\\ \hline
4	&4	&8	&1	&1	&0	&15.04	&0.63	&23.95	&38\\ \hline
\end{tabular}
\label{tbl_overlap}
\end{table}

The combination one has the highest speedup 27.82 and iteration 44. Its acceleration performance is the best but convergence performance is the worst. The combination four has an opposite effect with both the outer overlap and inner overlap set to 1. Its speedup is 23.95 and iteration is 38. If only the outer overlap or the inner overlap is set to 1, the results have an intermediate effect.

\subsection{AMG}

Two matrices, $ecology2$ and $3D\_Poisson$, are employed in the AMG algorithm testing. The $ecology2$ is a positive definite matrix derived from a circuit theory applied to animal/gene flow. It has 999,999 rows and 2,997,995 nonzero entries. The $NNZ/N$ is 3. The $3D\_Poisson$ has an dimension of 125,000 ($50\times50\times50)$) and 860,000 nonzero entries. Its $NNZ/N$ is about 7. We set the maximal level to 8 and the pre-smoothing and post-smoothing both to 3. The V-cycle is employed. Table~\ref{tbl_amg_ecology2} and \ref{tbl_amg_3dpoisson} collect the running results for $ecology2$ and $3D\_Poisson$, respectively. Two type of coarsening strategies Ruge- St\"{u}ben (RS) and CLJP are used. Two types of interpolations, the standard RS (RSSTD) and direct (RSD), are used. Four types of smoothers are tested. They are the damped Jacobi (dJacobi), weighted Jacobi (wJacobi), Chebyshev polynomial smoothers (Chev) and Gauss-Seidel (GS).

\begin{table}
\caption{AMG for \texttt{ecology2}}
\begin{tabular}{|r|r|r|r|r|r|r|r|} \hline

\bfseries Seq & \bfseries Coarsening & \bfseries               & \bfseries          & \bfseries CPU time & \bfseries GPU time & \bfseries         & \bfseries           \\
\bfseries No. & \bfseries strategy   & \bfseries Interpolation & \bfseries Smoother & \bfseries (second) & \bfseries (second) & \bfseries Speedup & \bfseries Iteration \\ \hline
1	          &CLJP	       &RSD	    &dJacobi	    &1.30	       &0.17	       &7.57	   &3\\ \hline
2	          &CLJP	       &RSD	    &Chev	        &4.92	       &0.50	       &9.78	   &11\\ \hline
3	          &RS	       &RSD	    &dJacobi	    &0.82	       &0.11	       &7.71	   &3\\ \hline
4	          &RS	       &RSSTD	 &wJacobi	    &0.86	       &0.12	       &7.07	   &3\\ \hline
5	          &RS	       &RSSTD	 &GS	        &0.46	       &0.99	       &0.46	   &1\\ \hline
\end{tabular}
\label{tbl_amg_ecology2}
\end{table}

The dJacobi, wJacobi and Chev are all developed based on the SpMV and vector operations. As we have completed the favorable parallel realization of them, these smoothers have good speedup for $ecology2$ , which are over 7. When the CLJP and RSD are used, the speedup reaches to the maximal value 9.78. If we select the GS smoother, the speedup is 0.46 which is very low. That means the running time on a GPU is even longer than that on a CPU. The purpose of acceleration on a GPU fails. Although the GS has the worst parallel performance, it has the best convergence performance and only once iteration is needed. So it is better to develop an AMG algorithm with the GS on a CPU. This also shows there is a contradictory effect between acceleration and convergence performance. Our experiment results show that the dJacobi, wJacobi and Chev are suitable for GPU computation while the GS is suitable for CPU.

\begin{table}
\caption{AMG for \texttt{3D\_Poisson}}
\begin{tabular}{|r|r|r|r|r|r|r|r|} \hline

\bfseries Seq & \bfseries Coarsening & \bfseries               & \bfseries          & \bfseries CPU time & \bfseries GPU time & \bfseries         & \bfseries           \\
\bfseries No. & \bfseries strategy   & \bfseries Interpolation & \bfseries Smoother & \bfseries (second) & \bfseries (second) & \bfseries Speedup & \bfseries Iteration \\ \hline
1	          &CLJP	   &RSD	        &dJacobi	    &1.64	       &0.65	       &2.54	       &8\\ \hline
2	          &CLJP	   &RSD	        &Chev	        &1.86	       &1.13	       &1.64	       &8\\ \hline
3	          &RS	   &RSD	        &dJacobi	    &0.25	       &0.05	       &4.71	       &7\\ \hline
4	          &RS	   &RSSTD	    &wJacobi	    &0.46	       &0.13	       &3.61	       &9\\ \hline
5	          &RS	   &RSSTD	    &GS	            &0.28	       &4.53	       &0.06	       &4\\ \hline
\end{tabular}
\label{tbl_amg_3dpoisson}
\end{table}

The $3D\_Poisson$ has worse acceleration results than the $ecology2$ has; shown by Table~\ref{tbl_amg_3dpoisson}. Different matrices has different nonzero patterns which have great influence on the computing performance. The algorithm on GPU has an acceleration effect for the smoothers of dJacobi, wJacobi and Chev. The combination three with the RS, RSD, dJacobi has the highest speedup 4.71. However, a very poor speedup 0.06 is obtained for the smoother GS. This result is similar to that of the matrix $ecology2$. The GS is not suitable for GPU computing is proved again.

\section{Conclusion}
\label{sec_conclusion}

We have developed the Krylov and AMG linear solvers on GPUs. The SpMV algorithm can be accelerated over 10 times faster on a single GPU against a CPU for most large-scale sparse matrices. Our preconditioned Krylov subspace algorithms have favorable speedups on GPUs. When four GPUs are employed and the inner RAS is set to 1024, the BiCGSTAB with ILU(0) can be sped up to 28 times faster. Our AMG solver shows good parallel performance for dJacobi, wJacobi and Chev smoothers. The numerical experiments verify that a contradictory effect exists between the performance of convergence and acceleration in many cases.

\subsubsection*{Acknowledgments.}
The support of Department of Chemical and Petroleum Engineering, University of Calgary and Reservoir Simulation Research Group is gratefully acknowledged. The research is partly supported by NSERC/AIEES/Foundation CMG, AITF iCore, IBM Thomas J. Watson Research Center, and the Frank and Sarah Meyer FCMG Collaboration Centre for Visualization and Simulation. The research is also enabled in part by support provided by WestGrid (www.westgrid.ca) and Compute Canada Calcul Canada (www.computecanada.ca).

\end{document}